\documentclass[twocolumn,floatfix,amsmath,amsfonts]{revtex4}
\usepackage{amsmath}
\usepackage{epsfig}

\begin{document}

\title{$n$-body dynamics of stabilized vector solitons.}

\author{Gaspar D. \surname{Montesinos}}
\affiliation{Departamento de Matem\'aticas, Escuela T\'ecnica
Superior de Ingenieros Industriales,\ \\ Universidad de Castilla-La
Mancha, 13071 Ciudad Real, Spain}

\author{Mar\'{\i}a I. \surname{Rodas-Verde}}
\affiliation{\'Area de \'Optica, Facultade de Ciencias de Ourense,\ \\
Universidade de Vigo, As Lagoas s/n, Ourense, ES-32005 Spain.}

\author{V\'{\i}ctor M. \surname{P\'erez-Garc\'{\i}a}}
\affiliation{Departamento de Matem\'aticas, Escuela T\'ecnica
Superior de Ingenieros Industriales, \\
Universidad de Castilla-La Mancha, 13071 Ciudad Real, Spain.}

\author{Humberto \surname{Michinel}}
\affiliation{\'Area de \'Optica, Facultade de Ciencias de Ourense,\\
Universidade de Vigo, As Lagoas s/n, Ourense, ES-32005 Spain.}

\date{November 5, 2004}


\begin{abstract}
In this work we study the interactions between 
 stabilized Townes solitons. By means of effective Lagrangian 
 methods, we have found that the interactions between
these solitons are governed by central forces, in a
first approximation. In our numerical
simulations we describe different types of orbits, 
deflections, trapping and soliton splitting. Splitting phenomena 
are also described by finite-dimensional reduced models.
All these effects could be used for potential
applications of stabilized solitons.
\end{abstract}


\maketitle

\textbf{The cubic Nonlinear Schr\"{o}dinger Equation (NLSE) is
among the most important physical models in the field of nonlinear
waves. Besides its fundamental value as a first order nonlinear wave
equation, it is an integrable model in the one dimensional case
\cite{Zakharov} and represents many different physical systems:
from laser wavepackets propagating in nonlinear materials to matter waves
in Bose-Einstein Condensates (BEC), gravitational models for quantum
mechanics, plasma physics or wave propagation in geological systems,
between others \cite{Akh,VVz,Sulem}. In this paper we consider the vector version of the NLSE and study the dynamics of some particular solutions which are stabilized by means of a periodic modulation of the nonlinearity.}


\section{Introduction}
\label{intro} 

One of the types of NLS equations arising frequently in the applications is the 
 two-dimensional cubic Nonlinear Schr\"odinger equation, which is of the form 
\begin{subequations}
\begin{eqnarray}
  \label{eq:gpe3}
  i \frac{\partial u}{\partial t} & = &
  \left[-\frac{1}{2}\triangle + g(t)|u|^2\right]u \\
  u(\boldsymbol{r},0) & = & u_0(\boldsymbol{r}) \in H^1(\mathbb{R}^2)
\end{eqnarray}
\end{subequations}
where $u(\boldsymbol{r},t): \mathbb{R}^2\times \mathbb{R}^+\rightarrow
\mathbb{C}$ is the complex wave amplitude, $\triangle =
\partial^2/\partial x^2 + \partial^2/\partial y^2$ and
$g(t)$ is a real function (the nonlinear coefficient) so that if
$g<0$ the nonlinearity is attractive whereas for $g>0$ the
nonlinearity is repulsive.

When $g$ is a real constant Eq. \eqref{eq:gpe3} is the cubic NLSE,
which is one of the most important models of mathematical physics.

It is well known that for $g<0$ if $N = \int_{\mathbb{R}^n} |u_0|^2$, is above
a threshold value $N_c$, solutions of Eq. \eqref{eq:gpe3} can
self-focus and become singular in a finite time. This phenomenon
is called \emph{wave collapse} or \emph{blowup of the wave
amplitude}. More precisely, there is never blowup when $N<N_c$ but
for any $\epsilon > 0$, there exist solutions with
$N=N_c+\epsilon$ for which blowup takes place
\cite{Fibich,Weinstein}.

Eq. \eqref{eq:gpe3} admits stationary solutions of the form
$u(\boldsymbol{r},t)=e^{i\mu t} \Phi_{\mu}(\boldsymbol{r})$, where $\Phi_{\mu} (\boldsymbol{r})$ verifies
\begin{equation}\label{soliton}
\triangle \Phi_{\mu} -2\mu \Phi_{\mu}- 2 g |\Phi_{\mu}|^2
\Phi_{\mu}=0.
\end{equation}
As it is precisely stated in \cite{Sulem}, when $g$ is negative,
for each positive $\mu$ there exists only one solution of Eq.
\eqref{soliton} which is real, positive and radially symmetric and
for which $\int |\Phi_{\mu}|^2 d\boldsymbol{r}$ has the minimum value between
all of the possible solutions of Eq. \eqref{soliton}. Moreover,
the positivity of $\mu$ ensures that this solution decays
exponentially at infinity. This solution is called the
\emph{ground state} or \emph{Townes soliton}. We will denote it as $R_{\mu}(r)$ which satisfies
\begin{eqnarray}\label{solitonR}
\triangle R_{\mu} -2\mu R_{\mu}- 2 g R_{\mu}^3=0 \\
\label{boundary} \lim_{r\rightarrow{\infty}} R_{\mu}(r)=0,\
R_{\mu}'(0)=0.
\end{eqnarray}

From the theory of nonlinear Schr\"{o}dinger equations it is known
that the Townes soliton has exactly the critical norm for blowup
$N_c$, therefore, it separates in some sense the region of
collapsing and expanding solutions. Moreover, the Townes soliton
is \emph{unstable}, i.e. small perturbations of this solution lead
to either expansion of the initial data or blowup in finite time. 
This instability is an essential feature of these type of equations which has its origin in the two-dimensional nature of the equation and makes it essentially different from its one-dimensional version, in which 
stable solitary-wave solutions exist, the so-called solitons.

From the physical point of view the existence of this instability implies that no localized solutions of the 2D NLSE exist. This is why there has been a great interest on the  case where $g$ is not constant but
a continuous periodic function of $t$, which has arisen recently in
different fields of applications of Eq. (\ref{eq:gpe3}). The
intuitive idea is that (oscillating) bound states could be obtained
by combining cicles of positive and negative $g$ values so that
after an expansion and contraction regime the solution could come
back to the initial state. In this way some sort of pulsating
trapped solution, i.e., a \emph{breather}, could be obtained.

This idea was first proposed in the field of nonlinear Optics
\cite{Berge}. In that context, a spatial modulation of the Kerr coefficient (the
nonlinearity) of the optical material is used to prevent collapse
so that the wavepacket becomes collapsing and expanding in alternating
regions and is stabilized in average \cite{Malomed,Malomed2}. The same idea has been used
in the field of matter waves in Refs. \cite{pisaUeda,pisaBoris}.
In Ref. \cite{Gaspar} some general results are provided for
generic forms of $g(t)$. Also in Refs. \cite{Gaspar,IMACS} it has been argued that the structure which remains stabilized is a Townes soliton. The stabilization of more complex structures in the framework of Eq. \eqref{eq:gpe3} constitutes an open problem 
since other solutions such as those of vortex type cannot be stabilized \cite{Malomed,montesinos04b}.

\section{THE VECTOR NLS AND STABILIZED VECTOR SOLITONS}

In this paper we explore the vector version of Eq. \eqref{eq:gpe3}, which is of the form
\begin{equation}
\label{Manakoveqs} i \frac{\partial u_j}{\partial t} = -{1 \over 2}
\nabla^2 u_j + g(t) \left(a_{j1} |u_1|^2+ \ldots +a_{jn} |u_n|^2
\right)u_j,
 \end{equation}
where $j = 1,\ldots, n$, $u_j$ are the complex amplitudes $\Delta =
\partial^2/\partial x^2 + \partial^2/\partial y^2, a_{jk} \in
\mathbb{R}$ are the nonlinear coupling coefficients and $g(t)$ is
a periodic function accounting for the modulation of the
nonlinearity. 

Eqs. (\ref{Manakoveqs}) are the natural extension of the Manakov
system \cite{Manakov} to two transverse dimensions and an arbitrary
number of components. In Optics, for spatial solitons, $t$ plays the role of the
propagation coordinate and $u_j$ are $n$ mutually incoherent laser
wavepackets. One-dimensional Manakov-type models have been extensively
studied in nonlinear Optics, mainly due to the potential
applications of Manakov solitons in the design of all-optical
computing devices \cite{todos}. In BEC these equations (with an
additional trapping term) describe the dynamics of multicomponent
two-dimensional condensates, $u_{j}$ being the wavefunctions for
each of the atomic species involved \cite{PRLdual,Nature}.

 Some features of this model have been described in Ref. \cite{Humberto}. In particular it is clear that if $u_j(\boldsymbol{r},0) = R_{\mu}(\|\boldsymbol{r}-\boldsymbol{r}_j\|)$, where $R_{\mu}$ is a \emph{Stabilized Townes Soliton (STS)}, and 
 the centers of the distributions are much more separated than the width of the Townes solitons, then we may have states with STS on each component. 
 Because of the Galilean invariance we can also construct solutions  propagating with uniform velocity along straight trayectories $\boldsymbol{r}_j(t)$. Again, if the trayectories are well separated it is reasonable to expect that (as it happens in the case of generic time-independent nonlinearities \cite{rig1,rig2})
 the STS will propagate without interactions. The purpouse of this paper is to provide a first systematic exploration of collisions of STS. A few results were already reported in Ref. \cite{Humberto}. One of the main contributions of that paper was to realize that 
for a given set of
parameters $a_{jk}$ it is possible to use STS to build explicit solutions of Eqs. \eqref{Manakoveqs}.
These solutions are constructed by taking
\begin{equation}
u_j= \alpha_jR_{\mu}(r), j=1,\ldots, n
\end{equation}
 for any set of coefficients $\alpha_j$ satisfying
\begin{equation}
\label{coefs} a_{j1} \alpha_1^2 + ... + a_{jn}\alpha_n^2 = 1,
j=1,\ldots, n.
\end{equation}
These solutions of Eqs. \eqref{Manakoveqs} are called \emph{Stabilized Vector Solitons (SVS)} because they correspond to solutions with appreciable overlapping of the different components $u_j$. In Ref. \cite{Humberto} the stability of these structures as well as the way they arise from collisions between stabilized solitons was studied.

In this paper we present many more examples of collisions between STS and the associated phenomenology.


\section{EFFECTIVE-PARTICLE MODEL FOR COLLISIONS OF STABILIZED TOWNES SOLITONS}
\label{theory}

\subsection{Motivation}
\label{motivation}

Before describing the direct numerical simulations of Eqs. \eqref{Manakoveqs} in detail and the many different phenomenologies observed we first present an effective-particle model for collisions of STS. This model will give us a few hints on the expected dynamics of the system. The idea, as in many other problems in Physics, is to assume that during the collisions stabilized solitons behave as particles in the sense that can be described qualitatively by a bell-type ansatz with a few free parameters. 

This type of assumptions allows to reduce the dynamics to a finite number degrees of freedom and receives many different names depending on the field of application: method of collective coordinates, 
averaged Lagrangian description, time-dependent variational method, effective-particle method, etc.

In our case we take as trial Gaussian functions, which is one of the standard choices
\begin{eqnarray}
\label{trial} u_{j}& = &
A_{j}\exp\left[-\frac{\left(x-x_{j}\right)^{2}}{2w_{jx}^{2}}
-\frac{\left(y-y_{j}\right)^{2}}{2w_{jy}^{2}}+\right. \\ \nonumber
& & \left.
+i\left(v_{jx}x+v_{jy}y+\beta_{jx}x^{2}+\beta_{jy}y^2\right)\right].
\end{eqnarray}

The $t$-dependent parameters have the following meaning:  $A_j$ is
the wavefunction amplitude; $x_j, y_j$ are the coordinates of the
centroid; $w_{jx}, w_{jy}$ are the
widths along the $x$ and $y$ axis; $v_{jx}, v_{jy}$ 
the initial velocities and $\beta_{jx},\beta_{jy}$ are phase factors which are required to obtain reliable results \cite{PhysicaD}.
Although Gaussians do not have the same asymptotic decay as STS,
our choice simplifies the calculations while leading to the same qualitative relations for the parameters and, as we will see below, the resulting equations provide an elegant and simple picture for the dynamics of the centroids of the STS.

\subsection{Equations for the effective-particle parameters}

Eqs. \eqref{Manakoveqs} can be derived by means of  a variational formalism 
from a Lagrangian density
which can be written as
a sum  of the Lagrangians for the linear operators plus the
nonlinear interaction in the following simple form:
\begin{subequations}
\label{L}
\begin{eqnarray}
\mathcal{L}&=&\frac{1}{2}\sum_{j,k=1}^{n}\left(\mathcal{L}_{j}+\mathcal{L}_{jk}\right), \\
\mathcal{L}_{j}&=& i\left(u_j\dot{u}_j^{\ast}-u_j^{\ast}\dot{u}_j\right)
+\left|\vec{\nabla}u_{j}\right|^{2},\\
\mathcal{L}_{jk}&=& g\left(t\right)a_{jk}\left|u_{j}\right|^{2}\left|u_{k}\right|^{2}.
\end{eqnarray}
\end{subequations}
The standard calculations of the method \cite{Boris} consist of
minimizing the trial functions from Eqs. \eqref{trial} over the
Lagrangian density given by Eqs. \eqref{L}. The final result is a
set of second order ordinary differential equations for the
evolution of the effective-particle parameters. In the case of the centroids
we obtain:
\begin{subequations}
\begin{eqnarray}
\label{centroid}
\ddot{x}_{j}&=&-\sum_{k=1}^{n}\frac{\partial I_{jk}}{\partial x_{j}},\\
\ddot{y}_{j}&=&-\sum_{k=1}^{n}\frac{\partial I_{jk}}{\partial y_{j}}.
\end{eqnarray}
\end{subequations}
With $j \neq k$ these equations are in the form of Newton's second
law with $I_{jk}$ playing the role of a potential ruling the
interaction between pairs of solitons according to the expression:
\begin{equation}
 I_{jk}=\frac{a_{jk}N_kg(t)}{\pi}\frac{
e^{-\frac{(x_{k}-x_{j})^{2}}{w_{jx}^{2}+w_{kx}^{2}}}}{\sqrt{w_{jx}^{2}
+w_{kx}^{2}}}\frac{e^{-\frac{(y_{k}-y_{j})^{2}}{w_{jy}^{2}
+w_{ky}^{2}}}}{\sqrt{w_{jy}^{2}+w_{ky}^{2}}}, \label{Ijk}
\end{equation}
being $N_k=\int|u_k|^2dxdy$ the square norm of the $k$-th wavepacket. The
corresponding widths along $x$ and $y$ are determined by the following equations\begin{subequations}
\begin{eqnarray}
\label{w} \ddot{w}_{jx}&=&\frac{1}{w_{jx}^{3}}
-\sum_{k=1}^{n}\frac{\partial I_{jk}}{\partial w_{jx}},\\
\ddot{w}_{jy}&=&\frac{1}{w_{jy}^{3}}
-\sum_{k=1}^{n}\frac{\partial I_{jk}}{\partial w_{jy}}.
\end{eqnarray}
\end{subequations}
Finally, some complementary relationships (first integrals of
motion) for the velocities and phase coefficients are also obtained
\begin{subequations}
\label{otras}
\begin{eqnarray}
v_{jx}&=&\dot{x}_{j}-2x_{j}\beta_{jx}\label{vjx},\\
v_{jy}&=&\dot{y}_{j}-2y_{j}\beta_{jy}\label{vjy},\\
\beta_{jx}&=&\frac{\dot{w}_{jx}}{2w_{jx}},\\
\beta_{jy}&=&\frac{\dot{w}_{jy}}{2w_{jy}}.
\end{eqnarray}\label{complem}
\end{subequations}
From Eqs. \eqref{complem} it is evident that the quantities $
v_{jx},v_{jy}$ play the role of the initial velocities of the 
distribution centroids.

\subsection{Fast modulation approximation} \label{approx}

In this paper we take the modulation $g(t)=g_0+g_1\cos\Omega t$ although the same qualitative results are obtained with other choices for $g$. To get STS the period of $g(t)$ must be very short \cite{Berge,Gaspar,IMACS} and the mean value $<g>$ must satisfy the relation $<g><-N_c$, where $N_c$ is the critical square norm for blowup. The oscillations of the
wavepacket widths, induced by $g(t)$, are very fast compared with the
dynamics of the centroids ruled by the potential from Eq.
\eqref{Ijk} with an effective range of the order of the sizes of the wavepackets.
Therefore, as a first approximation, the evolution of $x_j$ and
$y_j$ can be decoupled from the oscillations of $\omega_{jx}$ and
$\omega_{jy}$. With these assumptions the interaction between
different wavepackets will be determined by a constant nonlinearity with $g=<g>=g_0$.

\begin{figure}
\epsfig{file=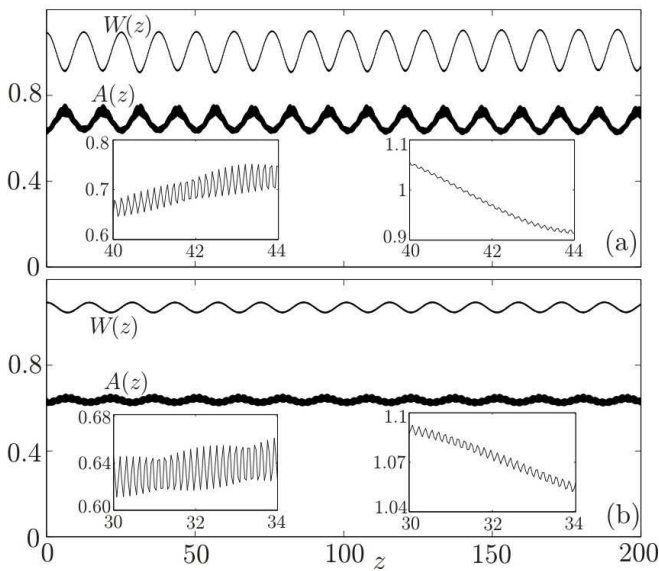,width=\columnwidth} \caption{Evolution of the
width $W=(\int (x^2+y^2)|u|^2)^{1/2}$ and of the peak amplitude $A=\max|u|$ of the numerical solution of
Eqs. \eqref{Manakoveqs} for $n=1$ with a nonlinear coefficient (a)
$g(t)=-2\pi+8\pi\cos(40t)$, (b) $g(t)=-2\pi+9.5\pi\cos(40 t)$.
Insets show details of the fast oscillations for the amplitude (left
panel) and for the width (right panel). \label{rapido}}
\end{figure}

On the other hand, the parameters of the modulation can be chosen
to minimize the amplitude of the wavepacket oscillations \cite{IMACS}. An example can be seen in Fig. \ref{rapido} where the results
of direct numerical integrations of Eqs. \eqref{Manakoveqs} in the
one-component case $n=1$ are shown. The peak amplitude and width
of the solution present a small variation when suitable parameters
are taken. Thus, as a first approximation 
we can consider wavepackets of constant
circular section $w_{jx}(t) = w_{jy}(t) = w_j$. With all these
considerations the potential between pairs of solitons is given by
\begin{equation}\label{Ijk2}
I_{jk}^0=\frac{a_{jk}N_kg_0}{\pi w^2} e^{-r_{jk}^2/w^2},
\end{equation}
where $w^2=w_j^2+w_k^2$ and $r_{jk}$ is the distance between
centroids.

Therefore, the interaction between two stabilized solitons is
governed, in a first approximation, by an attractive force which
only depends on the distance $r_{jk}$ between the centroids. That
is, the motion of the centroids will be similar to planetary mechanics, with a ``gravitational" potential $I_{jk}^0$ decaying as a Gaussian instead as
$1/r$. In fact, several common properties of central forces like
conservation of angular momentum and the reduction of the 2-body
problem to a single body motion in the center of mass frame are
straightforwardly derived. Specifically the reduced particle moves
under the action of an effective potential given by
\begin{equation}\label{potential}
V=\frac{a_{jk} N_k g_0}{\pi w^2}e^{-r^2/w^2}+\frac{J^2}{2r^2}
\end{equation}
where $g_0<0$ and $J$ is the conserved angular momentum.

Using elementary techniques it is easy to  obtain a necessary condition for the
existence of circular orbits, namely:
\begin{equation}\label{orbit}
C\Delta^2=e^{\Delta}
\end{equation}
where $C=2a_{jk}N_k|g_0|/\pi J^2$ is a constant of motion and
$\Delta=r^2/w^2$. If $C<e^2/4$ Eq. \eqref{orbit} has no solutions.
If $C=e^2/4$ there is only one ($\Delta_0=2$) and for $C>e^2/4$
there are two solutions ($\Delta_s<\Delta_0$ and
$\Delta_u>\Delta_0$). This means that depending on the initial
velocity of the wavepackets $v_0$ two closed orbits can exit. The most
internal one $(r_s<\sqrt{2}w)$ is stable whereas the external one
$(r_u>\sqrt{2}w)$ is unstable. If $v_0$ is increased up to a
threshold value $v_f$ (escape velocity) there exists only one
unstable closed orbit $(r_0=\sqrt{2}w)$. Finally if $v_0>v_f$ closed
orbits are not possible because the wavepackets move too fast to be
captured.

Taking this simple picture in mind, we will perform in the next
section a numerical exploration of the dynamics of stabilized solitons. Our main interest will be to check the existence of orbits and other dynamical processes.

\section{Numerical simulations of stabilized solitons collisions}
\label{simulations}

In this section we study the collisions of STS in the
framework of Eqs. \eqref{Manakoveqs}. We take initial data of the form 
\begin{equation}
u_j(\boldsymbol{r},0) = R_{0.5}(\|\boldsymbol{r}-\boldsymbol{r}_j\|) e^{i \boldsymbol{v}_j\cdot \boldsymbol{r}},
\end{equation}
and study their evolution by means of a pseudospectral Fourier scheme with time evolution of split-step type \cite{IMACS}. The scheme incorporates absorbing boundary conditions to get rid of the small amounts of radiation which are generated both by the STS and by the collisional processes.

Now, we proceed to describe the different behaviors observed.

\subsection{Fast collisions}

\begin{figure}
\epsfig{file=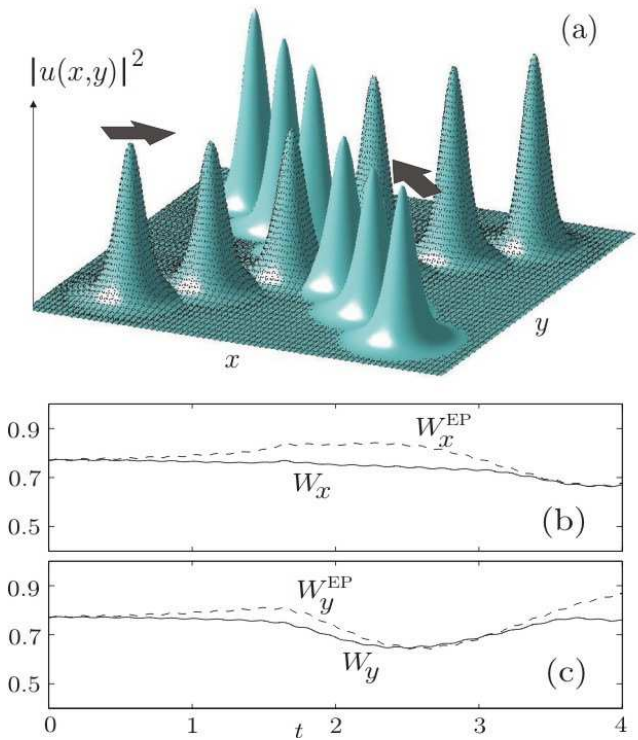,width=\columnwidth} \caption{[Color online] (a) Surface plots of $|u_1|^2$ and $|u_2|^2$ for times from $t=0$ to $t=3.4$ corresponding to the numerical simulation of Eqs. \eqref{Manakoveqs} with $g(t)=-2\pi+8\pi\cos(40 t)$, $\boldsymbol{r}_1=(-6,-6), \boldsymbol{r}_2=(6,-6), \boldsymbol{v}_1=(5/2^{1/2},5/2^{1/2})$ and $\boldsymbol{v}_2=(-5/2^{1/2},5/2^{1/2})$. (b) and (c) show the comparison of the evolution of the widths calculated numerically (solid line) and from the effective-particle model (dashed line) using Eqs. \eqref{variacional}.}\label{figdos}
\end{figure}

As it was shown in Ref. \cite{Humberto} when ``fast collisions" of STS take place the solitons emerge with only slight variations of the amplitudes and widths. These behaviors can be accounted for by the effective-particle model proposed in Sec. \ref{theory}. To see this we focus on the collisions of two stabilized solitions initially placed at $\boldsymbol{r}_1=(x_1,y_1), \boldsymbol{r}_2=(x_2,y_2)$ with initial velocities  $\boldsymbol{v}_1=(v_{1x},v_{1y}),\boldsymbol{v}_2=(v_{2x},v_{2y})$. Following the results of Sec. \ref{theory} and assuming trial Gaussian functions of equal size ($w_{1x}=w_{2x}=w_x$, $w_{1y}=w_{2y}=w_y$) we obtain the equation for the effective-particle parameters 
\begin{subequations}
\label{variacional}
\begin{eqnarray}
\ddot{x}_1 & = & -\frac{N g(t)}{2\pi w_x^3 w_y}e^{-\left(\frac{\ell_x^2}{2w_x^2}+\frac{\ell_y^2}{2w_y^2}\right)} \ell_x, \label{x1}\\
\ddot{x}_2 & = & -\ddot{x}_1,\\
\ddot{y}_1 & = & -\frac{N g(t)}{2\pi w_x w_y^3}e^{-\left(\frac{\ell_x^2}{2w_x^2}+\frac{\ell_y^2}{2w_y^2}\right)} \ell_y,\\ 
\ddot{y}_2 & = & -\ddot{y}_1,\\
\ddot{w}_x & = &
\frac{1}{w_x^3}+\frac{Ng(t)}{2\pi w_x^2 w_y}\left [
1+e^{-\left(\frac{\ell_x^2}{2w_x^2}+\frac{\ell_y^2}{2w_y^2}\right)}\left (
1-\frac{\ell_x^2}{w_x^2}\right)\right],\label{wa}\\
\ddot{w}_y & = & \frac{1}{w_y^3} + \frac{Ng(t)}{2\pi w_x
w_y^2}\left[1+e^{-\left(\frac{\ell_x^2}{2w_x^2}+\frac{\ell_y^2}{2w_y^2}\right)}\left (
1-\frac{\ell_y^2}{w_y^2}\right)\right] \label{wb},
\end{eqnarray}
\end{subequations}
where $\ell_x=x_2-x_1$ and $\ell_y=y_2-y_1$. Moreover we have the complementary relations \eqref{otras} and the conservation law $N(t) = \pi |A|^2 \omega_x \omega_y = \pi
|A(0)|^2 \omega_x(0) \omega_y(0)$. The different terms in Eqs.
\eqref{variacional} account for the phenomenology observed in ``fast collisions". For example,
they contain an asymmetric interaction (notice the differences
between Eqs. \eqref{wa} and \eqref{wb}) due to the fact that the width evolutions depend on the separation of both solitons along the corresponding axes. One example of this kind of interaction can be seen in Fig. \ref{figdos} where we plot the results of numerical simulation of Eqs. \eqref{Manakoveqs} with $\boldsymbol{r}_1=(-6,-6), \boldsymbol{r}_2=(6,-6), \boldsymbol{v}_1=(5/2^{1/2},5/2^{1/2}),\boldsymbol{v}_2=(-5/2^{1/2},5/2^{1/2})$. In this figure we also compare the results from Eqs. \eqref{Manakoveqs} with those obtained from Eqs. \eqref{variacional}. We can see that although there are quantitative differences the effective-particle model reproduces the observed dynamics and the qualitative behavior of the system.

\subsection{Collapsing orbits}

\begin{figure}
\epsfig{file=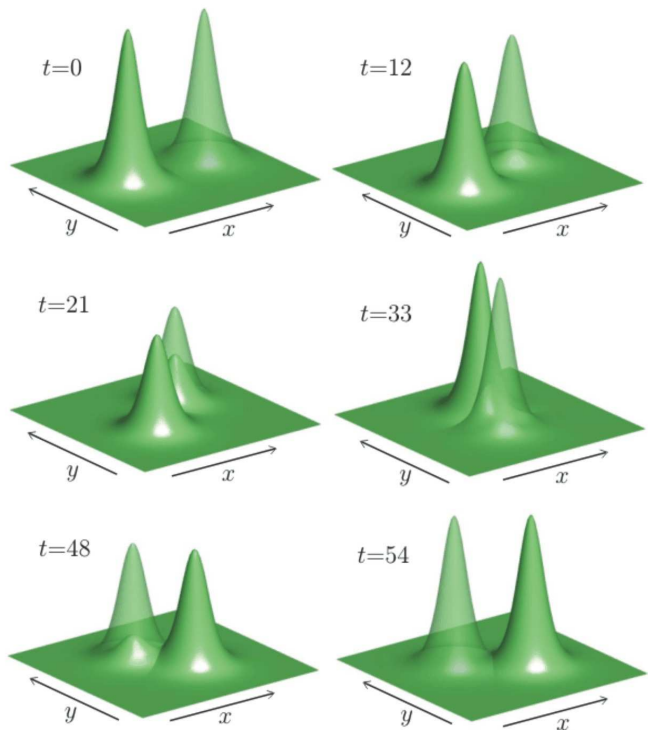,width=\columnwidth} \caption{[Color online] Surface
plots of $|u_1|^2$ (solid) and $|u_2|^2$ (transparent) for
different times corresponding to the simulation of
Eqs. \eqref{Manakoveqs} with $g(t)=-2\pi+8\pi\cos(40 t)$,
$\boldsymbol{r}_1=(-2,0)$, $\boldsymbol{r}_2=(2,0)$, $v_{1y}=-0.06$ and
$v_{2y}=0.06$.}\label{colapsante}
\end{figure}

We have studied the evolution of two stabilized solitons which
initially are placed at $\boldsymbol{r}_1=(-2,0)$ and $\boldsymbol{r}_2=(2,0)$ with initial
velocities $\boldsymbol{v}_{1y}=-0.06$ and $\boldsymbol{v}_{2y}=0.06$ along the $y$-axis. In
this situation we have observed that the solitons move inwards on
a spiraling orbit. Therefore, the distance between the centroids of the
two wavepackets
 decreases monotonically and the solitons join at the origin
periodically.

This behavior is shown  in Fig. \ref{colapsante} where we
plot the evolution of  $|u_1|^2$ and $|u_2|^2$ for different times. It can
be also observed that the solitons interact continuously and that
a partial splitting takes place, leading to the formation
of a pair of SVS \cite{Humberto}.

The effective-particle method cannot take into account the readjustment phenomenon between stabilized solitons 
observed in Fig. \ref{colapsante}. Therefore, this simplified description is not valid any more and can be used only as a first approximation to the problem. A more sophisticated model will be presented in Sec. \ref{variationalII}.

\subsection{Expanding orbits}

\begin{figure}
\epsfig{file=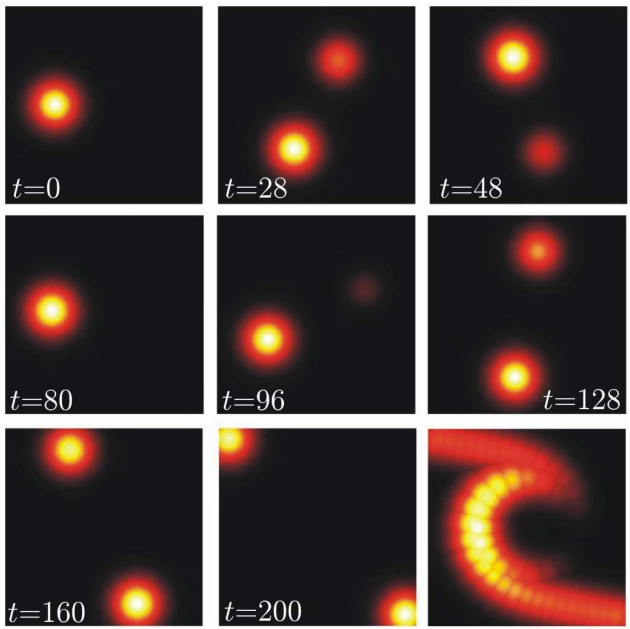,width=\columnwidth} \caption{[Color online] Pseudocolor
plots of $|u_1|^2$ for different times corresponding to the evolution of component $u_1$ from simulation
of Eqs. \eqref{Manakoveqs} with $g(t)=-2\pi+8\pi\cos(40 t)$,
$\boldsymbol{r}_1=(-2,0)$ and $v_{1y}=-0.08$. The last picture corresponds to
the superposition of several snapshots from $t=0$ to $t=200$.
\label{orbita0.08a}}
\end{figure}

\begin{figure}
\epsfig{file=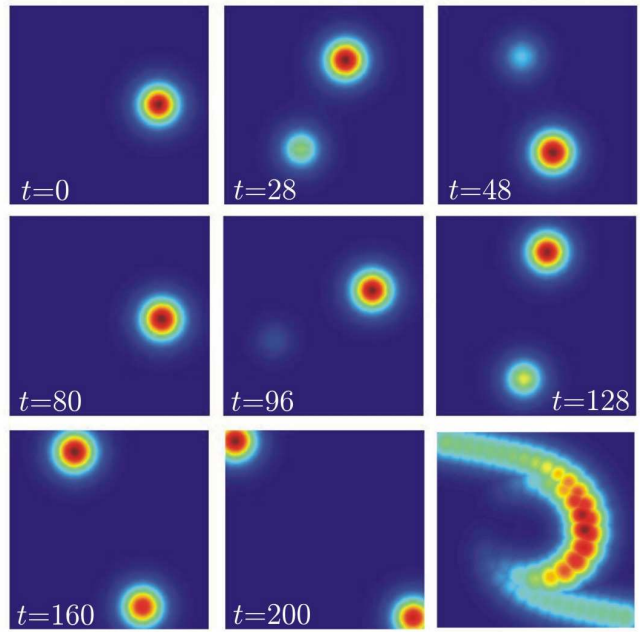,width=\columnwidth} \caption{[Color online] Same as Fig.
\ref{orbita0.08a} for the evolution of component $u_2$ with
$\boldsymbol{r}_2=(2,0)$ and $v_{2y}=0.08$.\label{orbita0.08b}}
\end{figure}

For initial velocities larger than in the previous case it is
possible to overcome the collapsing character of the orbit. If the initial
velocity is above a threshold value (which is analogous to a escape
velocity) the stabilized solitons follow outward trajectories. In Fig.
\ref{orbita0.08a} and Fig. \ref{orbita0.08b} we show the results
obtained for two stabilized solitons initially placed at
$\boldsymbol{r}_1=(-2,0)$ and $\boldsymbol{r}_2=(2,0)$ with input velocities $v_{1y}=-0.08$
and $v_{2y}=0.08$ along the $y$-axis.


Again we observe the formation of SVS. Since what it is plotted is
the amplitude of only one component ($|u_1|^2$ or $|u_2|^2$ ) the existence of two main spots in many subplots of Fig. \ref{orbita0.08a} and Fig. \ref{orbita0.08b} show, once more, that the interaction between stabilized solitons not only affects the trajectories of their centers, but also induces a readjustment of the distributions and the dynamical formation of SVS.
We will try to describe this phenomenon in the next section.

\subsection{Wavepacket splitting with deflection}

\begin{figure}
\epsfig{file=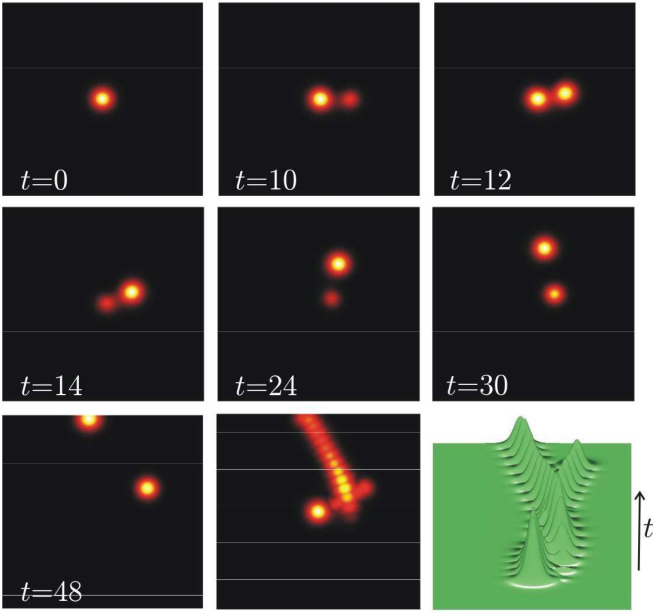,width=\columnwidth} \caption{[Color online] Pseudocolor
plots of $|u_1|^2$ for different times
corresponding to the evolution of component $u_1$ from simulation
of Eqs. \eqref{Manakoveqs} with $g(t)=-2\pi+8\pi\cos(40 t)$,
$\boldsymbol{r}_1=(0,0)$ and $\boldsymbol{v}_{1}=0$. The last pseudocolor plot corresponds
to the superposition of several snapshots from $t=0$ to $t=50$. It
is also shown as a surface plot in the bottom-right
picture.}\label{deflection1}
\end{figure}

\begin{figure}
\epsfig{file=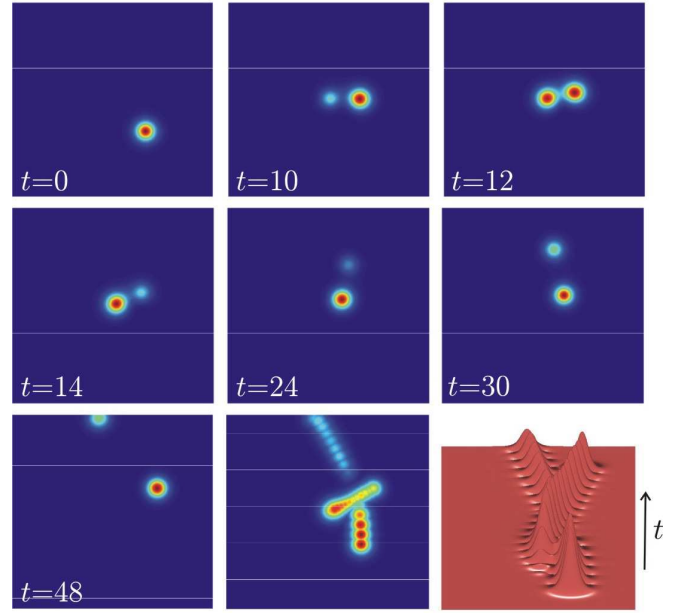,width=\columnwidth} \caption{[Color online] Same as Fig.
\ref{deflection1} for the evolution of component $u_2$ with
$\boldsymbol{r}_2=(3,-3)$ and $v_{2y}=0.3$.}\label{deflection2}
\end{figure}

In this subsection we consider 
another interesting case: one stabilized soliton initially
at rest and another one approaching to it. The
situation is shown in Fig. \ref{deflection1} and Fig.
\ref{deflection2} where we plot respectively the evolution of $u_1$
and $u_2$. The first wavepacket starts at $\boldsymbol{r}_1=(0,0)$ with zero initial
velocity and the second one is initially placed at $\boldsymbol{r}_2=(3,-3)$ with
initial velocity $v_{2y}=0.3$ along the $y$-axis. It can be
appreciated how the interaction leads to the formation of SVS by
means of the same splitting mechanism observed in previous figures. Pictures show that both wavepackets
split simultaneously and part of $u_1$ is dragged by one half
of $u_2$ forming a vector soliton. The remaining vector soliton
is deflected.


\subsection{Three interacting solitons}

\begin{figure}
\epsfig{file=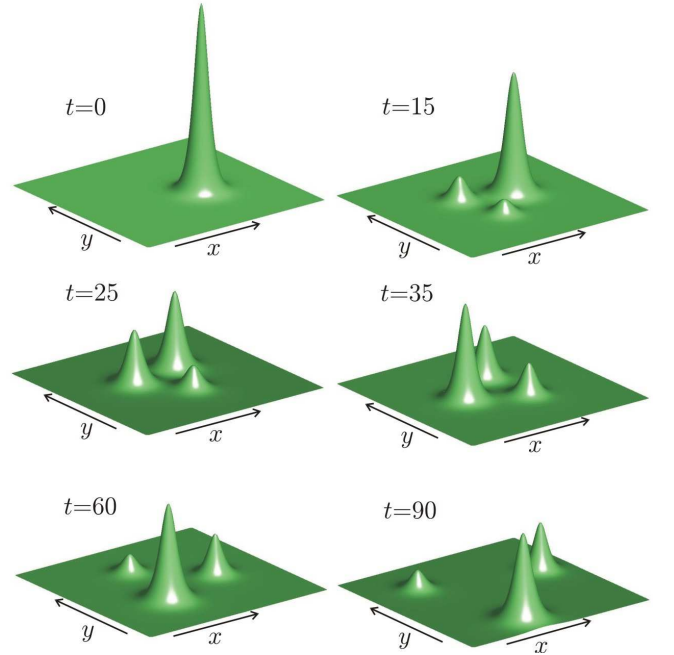,width=\columnwidth} \caption{[Color online] Surface plots
of $|u_3|^2$ for different times corresponding to
the evolution of component $u_3$ from simulation of Eqs.
\eqref{Manakoveqs} with $\boldsymbol{r}_1=(0,d/\sqrt{3})$,
$\boldsymbol{r}_2=(-d/2,-d/2\sqrt{3})$, $\boldsymbol{r}_3=(d/2,-d/2\sqrt{3})$ and initial
velocities $v_{1x}=-v\sin(\pi/2)$, $v_{1y}=0$,
$v_{2x}=-v\sin(\pi/2+2\pi/3)$, $v_{2y}=v\cos(\pi/2+2\pi/3)$,
$v_{3x}=-v\sin(\pi/2+4\pi/3)$, $v_{3y}=v\cos(\pi/2+4\pi/3)$ being
$d=4$, $v=0.12$ and $g(t)=-2\pi+8\pi\cos(40 t)$.}\label{triangle}
\end{figure}

We have also studied the three-body problem which corresponds to
three solitons in the corners of an equilateral triangle of side
$d$. Therefore, by taking the origin of coordinates in the
barycenter the solitons are placed at $\boldsymbol{r}_1=(0,d/\sqrt{3})$,
$\boldsymbol{r}_2=(-d/2,-d/2\sqrt{3})$, $\boldsymbol{r}_3=(d/2,-d/2\sqrt{3})$. The
initial velocities are $v_{1x}=-v\sin(\pi/2)$, $v_{1y}=0$,
$v_{2x}=-v\sin(\pi/2+2\pi/3)$, $v_{2y}=v\cos(\pi/2+2\pi/3)$,
$v_{3x}=-v\sin(\pi/2+4\pi/3)$ and $v_{3y}=v\cos(\pi/2+4\pi/3)$ being
$v$ the input velocity. The numerical results for $d=4$ and $v=0.12$
are shown in Fig. \ref{triangle} where we see the evolution of one
of the wavepackets. The initial velocities are such that the attraction
between solitons is not enough to keep them bounded and the wavepackets
move outwards. As in the previous cases the distributions split
during evolution, the initial distributions readjust due to the interaction
with the other two components and three two-part vector
solitons are generated.

\section{Theoretical analysis of the wavepacket splitting}
\label{variationalII}

\subsection{Motivation and ansatz}

From the previous section it is clear that the effective-particle model
used in Sec. \ref{theory} cannot explain the
wave\-packet splitting observed in the most of our numerical simulations. Therefore,
in this section we develop another approach to capture the main
characteristics of such splitting. We will use again an effective Lagrangian technique 
but now we  consider a two-mode model described by the following
ansatzs
\begin{subequations}
\label{trial2}
\begin{eqnarray}
u_{1}& = &
A_{11}\exp\left[-\frac{\left(x-\ell\right)^{2}}{2w_{1x}^{2}}-\frac{
y^{2}}{2w_{1y}^{2}}\right]+ \nonumber\\
& &+
A_{12}\exp\left[-\frac{\left(x+\ell\right)^{2}}{2\tilde{w}_{1x}^{2}}-\frac{
y^{2}}{2\tilde{w}_{1y}^{2}}\right],\\
u_{2}& = &
A_{21}\exp\left[-\frac{\left(x-\ell\right)^{2}}{2w_{2x}^{2}}-\frac{
y^{2}}{2w_{2y}^{2}}\right]+ \nonumber\\
& &
+A_{22}\exp\left[-\frac{\left(x+\ell\right)^{2}}{2\tilde{w}_{2x}^{2}}-\frac{
y^{2}}{2\tilde{w}_{2y}^{2}}\right].
\end{eqnarray}
\end{subequations}
This  choice now allows readjustment of mass between two parts                                                                                                 related to each component of the vector system, i.e. $u_1$ is now composed of two parts which can be used to emulate the splitting mechanism and the formation of SVS.

However, the full analysis of Eqs. (\ref{trial2}) by effective Lagrangian methods is cumbersome and to achieve some results we will make several
simplifications which can be justified by the observation of the
simulations. In the first place, when the initial stabilized solitons $u_1$ ans $u_2$ are equal, the splitting mechanism is a
symmetric process meaning that the growth rate of one part of
$u_1$ is the same that the growth rate of the corresponding part
of $u_2$. Therefore we can consider $A_{11}=A_{22}=\alpha$,
$A_{12}=A_{21}=\beta$, $w_{1x}=\tilde{w}_{2x}$,
$w_{1y}=\tilde{w}_{2y}$, $\tilde{w}_{1x}=w_{2x}$,
$\tilde{w}_{1y}=w_{2y}$. Secondly, we will assume that the sizes of every
part are very similar and remain close to their mean values except for the fast oscillations. So we consider that $w_{1x}=w_{2x}=\tilde{w}_{1x}=\tilde{w}_{2x}=w_x$ and
$w_{1y}=w_{2y}=\tilde{w}_{1y}=\tilde{w}_{2y}=w_y$ being $w_x, w_y$
constants. Finally, to get some insight on the problem we consider the case 
where the distance between
centroids $2\ell$ is approximately constant which can happen for instance during a fast switching process. Thus, the only free parameters are $\alpha$ and $\beta$.

\subsection{Model equations}

The Lagrangian $L=\int \mathcal{L} dxdy$ obtained from the density
Lagrangian $\mathcal{L}$ given by Eqs. \eqref{L} when the ansatzs
are as in Eqs. \eqref{trial2} is
\begin{eqnarray}
L &=&\pi w_x w_y\left\{i(\alpha\dot{\alpha}^*-\alpha^*\dot{\alpha}
+\beta\dot{\beta}^*-\beta^*\dot{\beta})+ \right.\nonumber\\
& + &i(\alpha\dot{\beta}^*-\alpha^*\dot{\beta}+\beta\dot{\alpha}^*-\beta^*\dot{\alpha})e^{-\frac{l^2}{w_x^2}}\nonumber+\\
& + &(|\alpha|^2+|\beta|^2)\left(\frac{1}{2 w_x^2}+\frac{1}{2 w_y^2}\right)+\nonumber\\
& + &(\alpha\beta^*+\beta\alpha^*)\left(\frac{w_x^2-2l^2}{2w_x^4}+\frac{1}{2w_y^2}\right)e^{-\frac{l^2}{w_x^2}}+\nonumber\\
& + & \frac{g(z)}{2}\left\{(|\alpha|^4+|\beta|^4)\left[1+e^{-\frac{2l^2}{w_x^2}}\right]+\right.\nonumber\\
& + & \left[6|\alpha|^2|\beta|^2+2(\alpha^2(\beta^*)^2+(\alpha^*)^2\beta^2)\right]e^{-\frac{2l^2}{w_x^2}}+\nonumber\\
& + &\left.\left.
 2|\alpha|^2|\beta|^2+4(|\alpha|^2+|\beta|^2)(\alpha\beta^*+\alpha^*\beta)e^{-\frac{3l^2}{2w_x^2}}\right\}\right\},
\end{eqnarray}
where we have taken $a_{11}=a_{12}=a_{21}=a_{22}=1$.

The standard calculations \cite{Boris} yield to a system
of ordinary differential equations for $\alpha$ and $\beta$,
\begin{eqnarray}
\label{system}
\begin{pmatrix}
1 & e^{-\frac{\ell^2}{w_x^2}}\\
e^{-\frac{\ell^2}{w_x^2}} & 1
\end{pmatrix}
\begin{pmatrix}
\dot{\alpha}\\
\dot{\beta}
\end{pmatrix}
& = &
-\frac{i}{2}
\begin{pmatrix}
\Lambda_1 & \Lambda_2\\
\Lambda_2 & \Lambda_1\\
\end{pmatrix}
\begin{pmatrix}
\alpha\\
\beta\\
\end{pmatrix},
\end{eqnarray}
where
\begin{subequations}
\begin{eqnarray}
\Lambda_1(t) & = &
\frac{1}{2w_x^2}+\frac{1}{2w_y^2}+ g(t)\left[\frac{2N(t)}{\pi w_x w_y} e^{-\frac{\ell^2}{2w_x^2}}+\right.\nonumber\\ 
& + & 
\left.(|\alpha|^2+|\beta|^2)(1+e^{-\frac{2\ell^2}{w_x^2}}-2e^{-\frac{\ell^2}{2w_x^2}})\right],\\
\Lambda_2(t) & = &  \left\{\frac{w_x^2-2\ell^2}{2w_x^4}+\frac{1}{2w_y^2} + g(t)\left[\frac{2N(t)}{\pi w_x w_y}+\right.\right.\nonumber\\ 
& + & 
\left.\left.(|\alpha|^2+|\beta|^2)(2e^{-\frac{\ell^2}{2w_x^2}}-2)\right]\right\} e^{-\frac{\ell^2}{w_x^2}},
\end{eqnarray}
\end{subequations}
and the function $N(t)$ is the square norm of the ansatzs
$$N(t)=||u_i||_2^2=\pi w_x w_y [|\alpha|^2+|\beta|^2+(\alpha\beta^*+ \alpha^*\beta)e^{-\frac{\ell^2}{w_x^2}}].$$
System \eqref{system} can be written as
\begin{eqnarray}
\label{systemdef}
\begin{pmatrix}
\dot{\alpha}\\
\dot{\beta}
\end{pmatrix}
& = & i
\begin{pmatrix}
\nu_1 & \nu_2\\
\nu_2 & \nu_1\\
\end{pmatrix}
\begin{pmatrix}
\alpha\\
\beta\\
\end{pmatrix},
\end{eqnarray}
where 
\begin{subequations}
\begin{eqnarray}
\nu_1(t) & = & \frac{\Lambda_2(t) e^{-\frac{\ell^2}{w_x^2}}-\Lambda_1(t)}{2(1-e^{-\frac{2\ell^2}{w_x^2}})},\\
\nu_2(t) & = & \frac{\Lambda_1(t) e^{-\frac{\ell^2}{w_x^2}}-\Lambda_2(t)}{2(1-e^{-\frac{2\ell^2}{w_x^2}})}	\label{nu2}.
\end{eqnarray}
\end{subequations}

Using system \eqref{systemdef} and its complex conjugate and taking into account that $\nu_1,\nu_2\in\mathbb{R}$ can be immediately proved that
\begin{subequations}
\begin{eqnarray}
\frac{d}{dt} (|\alpha|^2+|\beta|^2) & = & 0,\\
\frac{d}{dt} (\alpha\beta^*+\alpha^*\beta) & = & 0, 
\end{eqnarray}
\end{subequations}
and therefore three invariants exist
\begin{subequations}
\begin{eqnarray}
|\alpha|^2+|\beta|^2 & = & q_0,\\
\alpha\beta^*+\alpha^*\beta & = & q_1,\\
N(t) & = & N_0,
\end{eqnarray}
\end{subequations}
being $q_1,q_2,N_0$ constants.

System \eqref{systemdef} can be solved numerically to find the evolution of the amplitudes $\alpha$ and $\beta$. Nevertheless we can solve it analytically by considering that $\nu_1$ and $\nu_2$ are constants since the
only dependence on $t$ is given by the nonlinear coefficient $g(t)$
and, as we discussed in Sec. \ref{approx}, the dynamics is determined by an averaged nonlinearity $g=g_0$. By writing $\alpha=\alpha_R+i\alpha_I$, $\beta=\beta_R+i\beta_I$
with $\alpha_R$, $\alpha_I$, $\beta_R$, $\beta_I \in\mathbb{R}$ the solutions of system \eqref{systemdef} are obtained by using basic
techniques from the theory of ordinary differential equations and the result is
\begin{eqnarray}\label{solgen}
\begin{pmatrix}
\alpha_R\\
\beta_R\\
\alpha_I\\
\beta_I
\end{pmatrix}
& = & \frac{1}{2} M
\begin{pmatrix}
\sin(\nu_1-\nu_2)t\\
\cos(\nu_1-\nu_2)t\\
\sin(\nu_1+\nu_2)t\\
\cos(\nu_1+\nu_2)t\\
\end{pmatrix},
\end{eqnarray}
where
\begin{eqnarray}
M & = &
\begin{pmatrix}
-M_1 & M_2 & -M_3 & M_4\\
M_1 & -M_2 & -M_3 & M_4\\
M_2 & M_1 & M_4 & M_3\\
-M_2 & -M_1 & M_4 & M_3
\end{pmatrix},
\end{eqnarray}
and
\begin{subequations}
\begin{eqnarray}
M_1& = &
\alpha_{I}(0)-\beta_{I}(0),\\
M_2 & = & \alpha_{R}(0)-\beta_{R}(0),\\
M_3 & = &
\alpha_{I}(0)+\beta_{I}(0),\\
M_4& = & \alpha_{R}(0)+\beta_{R}(0),
\end{eqnarray}
\end{subequations}
being $\alpha_{R}(0), \alpha_{I}(0), \beta_{R}(0), \beta_{I}(0)$ the
initial conditions.

\subsection{Validation}

Taking as initial conditions $\alpha_R(0)=\alpha_0$,
$\alpha_I(0)=\beta_R(0)=\beta_I(0)=0$ from Eq. \eqref{solgen} we obtain
\begin{subequations}
\label{solution}
\begin{eqnarray}
\alpha_R & = & \alpha_0\cos \nu_1 t\cos \nu_2 t,\\
\beta_R & = & -\alpha_0\sin \nu_1 t\sin \nu_2 t,\\
\alpha_I & = & \alpha_0\sin \nu_1 t\cos \nu_2 t,\\
 \beta_I & = & \alpha_0\cos \nu_1 t\sin \nu_2 t,
\end{eqnarray}
\end{subequations}
and the amplitude evolutions are straightforwardly derived 
\begin{subequations}
\label{peak}
\begin{eqnarray}
|\alpha| & = & \alpha_0|\cos \nu_2 t|,\\
|\beta| & = & \alpha_0|\sin \nu_2 t|.
\end{eqnarray}
\end{subequations}

\begin{figure}
\epsfig{file=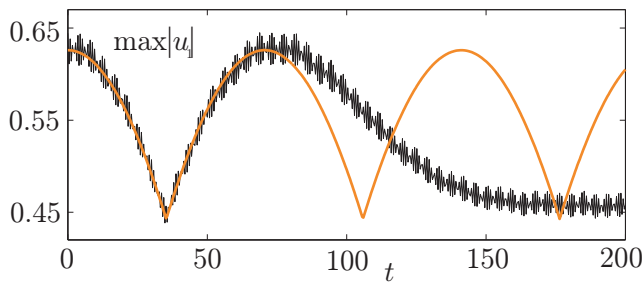,width=\columnwidth} \caption{Comparison
of the evolution of the peak amplitude of the wavepacket $u_1$ corresponding to the
simulation of Fig. \ref{orbita0.08a} and Fig. \ref{orbita0.08b}, obtained by direct numerical simulation of Eqs. \eqref{Manakoveqs} with the periodic oscillations from Eqs. \eqref{peak} predicted by the wavepacket splitting model.\label{varsplit}}
\end{figure}

Eqs. \eqref{peak} mean that the splitting mechanism is oscillatory
and periodic as can be seen in Figs.
\ref{colapsante}-\ref{deflection2}. In Fig. \ref{varsplit} we
compare the evolution of the peak amplitude of the wavepacket $u_1$ corresponding to the
simulation of Fig. \ref{orbita0.08a} and Fig. \ref{orbita0.08b}
(for which $\ell=2$), obtained by direct numerical simulation of Eqs.
\eqref{Manakoveqs} with Eqs. \eqref{peak}. The value of $\nu_2$ is calculated according to Eq. \eqref{nu2} with a slight correction in the value of $\ell$ which is taking as $\ell=1.88$. This correction is due to
the fact that numerical simulations of Fig. \ref{orbita0.08a} and Fig. \ref{orbita0.08b} are made with Townes solitons intial data whereas the analysis of the wavepacket splitting is made under the assumption of Gaussian initial data. Therefore, since Townes solitons decay at infinity as $\exp(-r)$ and Gaussians do it as $\exp(-r^2)$, for a fixed separation $2\ell$,  the overlapping between Townes solitons is greater than between Gaussians. For these reason to compare both situations it is necessary to take a smaller value of $\ell$ in the case of Gaussian data. We see that there is a very good agreement between both curves up to around $t=70$. After that numerical simulations show that solitons repel each other and the theoretical analysis is not valid any more, because we supposed constant separation between solitons. Therefore, we
can conclude that our two-mode model is valid to predict the
readjustment of mass between the wavepackets  provided that the distance between
parts remains nearly constant. As in the one dimensional case, the final repulsion of the wavepackets
can be explained by taking into account that the relative phases
between coherent solitons tend to separate them \cite{Kivshar}. In this case, the
solution of Eqs. \eqref{solution} gives a phase difference which
takes the values $\pm\pi/2$.

Finally, we must notice that the formation of vector solitons is the dominant tendency of the system in the slow collision regime. Therefore, the effective-particle model must be combined with the analysis of the partial splitting in order to get a more accurate picture of the dynamics. In fact this kind of solitons exhibit a teleportation behavior as they suddenly vanish to appear in another place. This effect is very fast and can have potential applications in optical information processing.


\section{Conclusions}
\label{conclusions}

In this paper we have presented a detailed study of interactions
between wavepackets which are stabilized against collapse by means of a
modulation of the nonlinear term, i.e. stabilized solitons. 

We have studied their dynamics by direct numerical simulations of the model equations. In this study,  we have found that the system
presents collapsing and expanding orbits depending on the initial
configuration as well as other phenomena as deflection of one wavepacket
due to the attraction and split in several parts in the
corresponding $n$-body interactions. 

We have also developed a
theoretical explanation of the wavepacket splitting based on 
effective Lagrangian methods and corroborated that effective-particle models allow to obtain conclusions for this system only in very limited situations.


G. D. M. and V. M. P-G. are partially supported by Ministerio de
Educaci\'on y Ciencia under grant BFM2003-02832 and Consejer\'{\i}a
de Educaci\'on y Ciencia of the Junta de Comunidades de Castilla-La
Mancha under grant PAC-02-002. G. D. M. acknowledges support from
grant AP2001-0535 from MECD.


\end{document}